\documentclass[useAMS,usenatbib,psfig]{mn2e}
\usepackage{amssymb}
\usepackage{psfig}
\usepackage{times}
\usepackage{color}
\usepackage{ulem}

\title[Limits on luminosity and accretion rate of a disc] {Limits on luminosity and mass accretion rate of a radiation pressure dominated accretion
disc}
\author[X. Cao \& W.-M. Gu]
{Xinwu Cao$^{1,2,3}$ and Wei-Min Gu$^3$\\
$^1$ Shanghai Astronomical Observatory, Chinese Academy of Sciences,
80 Nandan Road, Shanghai, 200030, China; E-mail: cxw@shao.ac.cn\\
$^2$ Key Laboratory of Radio Astronomy, Chinese Academy of Sciences,
210008 Nanjing, China\\
$^3$ Department of Astronomy and Institute of Theoretical Physics
and Astrophysics, Xiamen University, Xiamen, Fujian, 361005,
China;\\
E-mail: guwm@xmu.edu.cn}

\date{Accepted 2015 February 09. Received 2015 January 19; in original form 2014 June 29}

\pagerange{\pageref{firstpage}--\pageref{lastpage}} \pubyear{}

\begin{document}

\maketitle \label{firstpage}

\begin{abstract}
{There is a maximum for the gravity of a black hole in the vertical
direction in the accretion disc. Outflows may probably be driven
from the disc if the radiation flux of the disc is greater than a
critical value corresponding to the maximal vertical gravity.} We
find that outflows are driven by the radiation force from the disc
if the dimensionless mass accretion rate at the outer radius
$\dot{m}_{\rm out}\ga1$ ($\dot{m}=\dot{M}/\dot{M}_{\rm Edd}$,
$\dot{M}$ is the mass accretion rate, $\dot{M}_{\rm Edd}=L_{\rm
Edd}/0.1c^2$, {and $L_{\rm Edd}$ is the Eddington luminosity}).
{Assuming the outflow to be strong to carry away sufficient gas from
the disc surface,} we find that the radiation of the disc is limited
by such outflows. The disc luminosity, $L_{\rm disc}/L_{\rm
Edd}\propto\ln \dot{m}_{\rm out}$, at large-$\dot{m}_{\rm out}$
cases. The Eddington ratio of the disc is $\sim 3$ for $\dot{m}_{\rm
out}\sim 100$, which is significantly lower than that of a
conventional slim disc without outflows {\citep*[but it is
comparable with that given in][]{2003ApJ...593...69K}}. This implies
that the emission from some ultra-luminous X-ray sources with highly
super Eddington luminosity should be Doppler beamed, or intermediate
mass black holes are in these sources instead of stellar mass black
holes. The spectra of the discs surrounding massive black holes with
outflows are saturated in the {high frequency end} provided
$\dot{m}_{\rm out}\ga 2$. We suggest that the saturated emission can
be observed to estimate the masses of the black holes accreting at
high rates, such as the narrow-line Seyfert galaxies, with the model
calculations. The rate of the mass accreted by the black hole always
$\dot{M}_{\rm in}\simeq \dot{M}_{\rm Edd}$ even if the mass
accretion rate at the outer radius $\dot{M}_{\rm out}\gg
\dot{M}_{\rm Edd}$, because most of the gas is removed into the
outflows by the radiation force. If this is the case, the luminous
quasars at high redshifts $z\ga 6$ should have grown up through
persistent accretion at a rate close to the Eddington rate.
\end{abstract}


\section{Introduction}\label{intro}

Black hole accretion is the main energy source in different types of
the objects, such as, X-ray binaries, and active galactic nuclei
(AGN). In the past several decades, the measurements of masses of
black holes in AGN with the reverberation mapping method have been
widely used in the community
\citep*[e.g.,][]{1993PASP..105..247P,1999ApJ...526..579W,2000ApJ...533..631K,2004ApJ...613..682P}.
The Eddington ratio distributions of large AGN samples show that the
bolometric luminosity of luminous AGN is roughly Eddington limited
\citep[e.g.,][]{2006ApJ...648..128K}, or only a small fraction of
AGN are super-Eddington
\citep*[e.g.,][]{2002ApJ...579..530W,2004ApJ...608..136W,2009MNRAS.398.1905W,2012ApJ...746..169S,2012ApJ...761..143N,2013ApJ...764...45K,2014ApJ...787L..20D},
but also see the discussion in \citet{2014MNRAS.438..672N}. The main
observational features of AGN can be explained by accretion disc
models quite well. The standard thin accretion disc model has been
successfully used for the objects with moderate Eddington ratios,
$L/L_{\rm Edd}\la 0.3$ \citep*[$L$ is the bolometric luminosity and
$L_{\rm Edd}$ is the Eddington
luminosity,][]{1973A&A....24..337S,1989MNRAS.238..897L}. For more
luminous objects, the discs are no longer thin due to strong
radiation pressure, and slim accretion discs are suggested to be in
these objects \citep*[][]{1988ApJ...332..646A}. The slim disc models
have been extensively studied either analytically or with numerical
simulations
\citep*[e.g.,][]{1996ApJ...458..474S,1999ApJ...522..839W,2003A&A...398..927W,2007ApJ...660..541G,2010ApJ...715..623L,2011A&A...527A..17S,1988ApJ...330..142E,2005ApJ...628..368O,2007ApJ...670.1283O,2011ApJ...736....2O,2014ApJ...780...79Y}.
{The viscous timescale in the slim disc is shorter than the cooling
timescale in slim discs \citep*[][]{2003ApJ...593...69K}. A fraction
of photons are unable to escape from the disc, and therefore they
are trapped in the disc \citep*[][]{1978MNRAS.184...53B}. The
radiation efficiencies of slim discs are lower than those for
standard thin discs due to the radial energy advection in the disc
\citep*[][]{1996ApJ...458..474S,2000PASJ...52..133W,2006ApJ...648..523W}.
The detailed studies show that the luminosity of a slim disc
increases slowly with mass accretion rate as, $L/L_{\rm Edd}\propto
\ln\dot{m}$ ($\dot{m}=\dot{M}/\dot{M}_{\rm Edd}$, $\dot{M}$ is the
mass accretion rate, and $\dot{M}_{\rm Edd}=L_{\rm Edd}/0.1c^2$), at
high accretion rates, and the luminosity of the slim disc is almost
saturated at around $10 L_{\rm Edd}$
\citep*[][]{2000PASJ...52..133W,2006ApJ...648..523W}.}

Unlike most analytical works on slim discs, outflows are always
present in the numerical simulations of super-critical accretion
flows
\citep*[][]{2005ApJ...628..368O,2011ApJ...736....2O,2014ApJ...780...79Y}.
This implies that only a small fraction of gas in the disc is
finally swallowed by the black hole. \citet{2006ApJ...652..518M}
found that the gas will be blown away by the radiation force if the
relative half-thickness of the disc is $H/R>1/\sqrt{2}$ (see
equation 7 and the discussion in their paper), in which an exact
expression of the vertical component of gravity is adopted instead
of the approximation $GMz/R^3$ adopted in most literature, which is
only valid in the case of $H/R\ll 1$. A series of works on the
vertical structure of the slim disc indeed show that outflows are
inevitably driven from the the disc surface provided the luminosity
of the disc reaches the Eddington value
\citep*[][]{2007ApJ...660..541G,2009PASJ...61.1313G,2009ApJ...693..670J,2012ApJ...753..118G}.
Indeed, outflows have been observed in many luminous quasars
\citep*[e.g.,][]{2003MNRAS.346.1025P,2005ApJ...631L..33W,2006ApJ...639..716Z,2008ApJ...672..102G,2008ApJ...676L..97W,2009ApJ...702..851W,2010ApJ...719..700T,2011ApJ...742...44T,2012MNRAS.422L...1T}.

In this work, we investigate the structure of super-critical
accretion disc with outflows driven by the radiation force of the
disc. We describe our model calculations and results in Sects.
\ref{condt_outflow} and \ref{disc_outfl}. The last section contains
the discussion of the results.

\section{The condition for outflows driven by radiation
force}\label{condt_outflow}

It was suggested that the outflows are driven by the radiation force
at the disc surface when the relative disc thickness $H/R$ is
greater than a certain value
\citep*[][]{2006ApJ...652..518M,2007ApJ...660..541G}. In this
section, we extend the analysis on this issue in
\citet{2006ApJ...652..518M}.

The equilibrium in the vertical direction of an accretion disc gives
\begin{equation}
{\frac {{\rm d}p}{{\rm d}z}}=-{\frac {GM\rho z}{(R^2+z^2)^{3/2}}},
\label{vert_disc1}
\end{equation}
at radius $R$. For a radiation pressure dominated accretion disc,
\begin{equation}
p={\frac {\varepsilon_{\rm r}}{3}}, \label{p_r1}
\end{equation}
where $\varepsilon_{\rm r}$ is the energy density of radiation.
Substituting equation (\ref{p_r1}) into (\ref{vert_disc1}), we have
\begin{equation}
{\frac {1}{3}}{\frac {{\rm d}\varepsilon_{\rm r}}{{\rm d}z}}=-{\frac
{GM\rho z}{(R^2+z^2)^{3/2}}}=-{\frac {q(z)\kappa_{\rm T}\rho}{c}},
\label{dedz1}
\end{equation}
where $\kappa_{\rm T}$ is the electron scattering opacity, and $q$
is the radiation energy flux in $z$-direction. We obtain
\begin{equation}
q(z)={\frac {GMz}{(R^2+z^2)^{3/2}}}{\frac {c}{\kappa_{\rm T}}},
\label{q1}
\end{equation}
from equation (\ref{dedz1}).

The half-thickness $H$ of the disc is related to the outgoing
radiation flux $f_{\rm rad}=q(H)$ from the unit surface area of the
disc at $z=H$ by
\begin{equation}
f_{\rm rad}=q(H)={\frac {GMH}{(R^2+H^2)^{3/2}}}{\frac
{c}{\kappa_{\rm T}}}, \label{f_rad1}
\end{equation}
which can be re-written in dimensionless form,
\begin{equation}
\tilde{f}_{\rm rad}={\frac {\tilde{H}}{(1+\tilde{H}^2)^{3/2}}}.
\label{f_rad2}
\end{equation}
The dimensionless quantities are defined as
\begin{equation}
\tilde{f}_{\rm rad}=f_{\rm rad}{\frac {R^2\kappa_{\rm
T}}{GMc}},~~~~~{\rm and}~~~~~\tilde{H}={\frac
{H}{R}}.\label{dimensionless}
\end{equation}
{The estimate of $H$ in this work is different from and twice of
$H^\prime=c_{\rm s}/\Omega_{\rm K}$ in most previous works of
accretion discs (e.g., Frank, King \& Raine 2002; Kato, Fukue \&
Mineshige 2008; see Section 4.1 for the details).}

\begin{figure}
\centerline{\psfig{figure=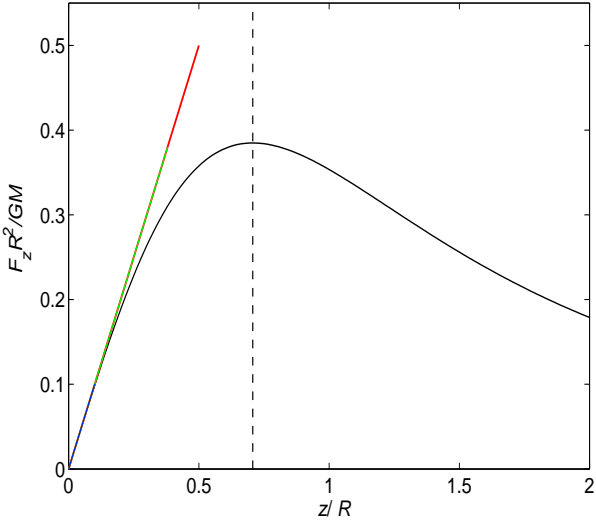,width=8.0cm,height=7.0cm}}
\caption{The vertical component of gravity as a function of $z$. The
dashed line corresponds $z/R=\sqrt{2}/2$. {The coloured lines
represent the radiation force in the case of the vertically constant
energy dissipation rate with different values of radiation flux,
$\tilde{f}_{\rm rad}=0.5$ (red), $2\sqrt{3}/9$ (green), and $0.1$
(blue).} } \label{f_grav_z}
\end{figure}

The vertical component of gravity of the black hole increases from
the mid-plane of the disc along $z$, and reaches the maximum at
$z/R=\sqrt{2}/2$ (see equation \ref{vert_disc1} and Fig.
\ref{f_grav_z}). The balance between the radiation force and the
vertical gravity is broken down if the radiation flux is
sufficiently large. This implies that outflows may probably be
driven from the disc if the radiation flux of the disc is greater
than a critical value $f_{\rm rad}^{\rm max}$ corresponding to the
maximal vertical gravity, which is given by
\begin{equation}
\tilde{f}_{\rm rad}^{\rm max}={\frac {\tilde{H}_{\rm
max}}{(1+\tilde{H}_{\rm max}^2)^{3/2}}}={\frac {2\sqrt{3}}{9}},
\label{f_radmax1}
\end{equation}
where $\tilde{H}_{\rm max}=\sqrt{2}/2$ is used.

{Both the vertical gravity and radiation force exerted on the gas in
the disc is proportional to gas density $\rho$, and therefore the
density $\rho$ is canceled in the vertical force balance equation
(\ref{dedz1}). We do not need to assume a specific function of $z$
for gas density in the above analysis. This means our conclusion
does not depend on the vertical density distribution of the disc.
\citet{2004ApJ...605L..45T} found that the energy dissipation is
roughly constant vertically in the radiation-MHD simulation for a
radiation-dominated accretion disc. If this is the case, the
radiation energy flux in the disc is
\begin{equation}
q(z)=f_{\rm rad}{\frac {z}{H}} \label{f_rad4}
\end{equation}
The radiation force exerted on the unit of mass in $z$-direction is
\begin{equation}
F(z)=q(z){\frac {\kappa_{\rm T}}{c}}=\tilde{f}_{\rm rad}{\frac
{z}{H}}{\frac {GM}{R^2}}. \label{force_rad1}
\end{equation}
We plot the radiation force in Fig. \ref{f_grav_z} to compare with
the vertical component of gravity. It is found that the radiation
force is almost in equilibrium with gravity in $z$-direction for the
low-$\tilde{f}_{\rm rad}$ (i.e., low-$\tilde{H}$) case. This is the
case considered in the radiation MHD simulation given in
\citet{2004ApJ...605L..45T}. The vertical gravity $-\rho\Omega_{\rm
K}^2z$ is adopted in \citet{2004ApJ...605L..45T}'s work, which is a
good approximation for $z/R\ll 1$. When $z/R$ is sufficiently large
($\ga 0.2$), the vertical gravity deviates obviously from the linear
relation of $z$ (see equation \ref{vert_disc1}). If the energy
dissipation can still remain constant vertically in this case, the
radiation force may be dominant over the vertical gravity at the
upper lay of the disc (see Fig. \ref{f_grav_z}). It should be
cautious on this conclusion, because the disc thickness adopted here
is derived on the assumption of the vertical gravity
$-\rho\Omega_{\rm K}^2z$. This issue is to be resolved by the future
numerical simulation adopting an exact function of vertical gravity.
In the case of $\tilde{f}_{\rm rad}>\tilde{f}_{\rm rad}^{\rm max}$,
the radiation force always dominates over the vertical gravity in
the region of $z/R>\tilde{H}_{\rm max}$ for arbitrary $z$-dependent
energy dissipation rate even if the disc thickness is not well
determined. This can be verified by shifting $\tilde{f}_{\rm rad}$
($\tilde{f}_{\rm rad}>\tilde{f}_{\rm rad}^{\rm max}$) in the
horizontal direction in Fig. \ref{f_grav_z} for arbitrary value of
$\tilde{H}$.}

{The condition of outflows driven by radiation force is derived
above for a radiation pressure dominated accretion disc, which is
probably the case for a disc accreting at a high rate. We should
point out that the condition is valid even if the disc is not
radiation pressure dominant. The condition for radiation pressure
driving outflows is the radiation force being greater than the
maximum of the vertical gravity, i.e.,
\begin{equation}
{\frac {f_{\rm rad}\rho\kappa_{\rm T}}{c}}>\rho {\cal F}_{z}^{\rm
max},
\end{equation}
where ${\cal F}_{z}$ is the vertical gravity for unit mass of gas,
\begin{equation}
{\cal F}_z(R,z)={\frac {GMz}{(R^2+z^2)^{3/2}}},
\end{equation}
of which the maximum is ${\cal F}_{z}^{\rm max}=2\sqrt{3}GM /9R^2$.
In dimensionless form, it becomes
\begin{equation}
\tilde{f}_{\rm rad}>2\sqrt{3}/9,
\end{equation}
where equation (\ref{dimensionless}) is used. This is exactly the
same as the condition derived in the first part of this section (see
equation \ref{f_radmax1}).} {The self-gravity of the disc may
dominate over the gravity of the black hole in the outer region of
the disc, which is not considered in this work. This may alter the
condition for outflows driven by radiation force, and the
self-gravity may hamper the disc to launch outflows. }


\section{The accretion disc with outflows}\label{disc_outfl}

For a disc accreting at a high rate, the outgoing radiation flux
$\tilde{f}_{\rm rad}>\tilde{f}_{\rm rad}^{\rm max}$. In this case, a
fraction of gas in the disc will be channeled into the outflows by
the radiation force. The mass accretion rate in the disc will be a
function of radius. Following the standard accretion disc model
\citep*[][]{1973A&A....24..337S,2002apa..book.....F}, we derive the
dissipation power in unit surface area of the disc with a
$R$-dependent mass accretion rate $\dot{M}(R)$,
\begin{equation}
D(R)={\frac {3GM\dot{M}(R)}{8\pi R^3}}\left[1-{\frac {\dot{M}_{\rm
in}}{\dot{M}(R)}}\left({\frac {R_{\rm in}}{R}}\right)^{1/2}\right],
\label{d1}
\end{equation}
where $\dot{M}_{\rm in}$ is the mass accretion rate at the inner
radius of the disc $R_{\rm in}$. For a non-rotating black hole,
$R_{\rm in}=6GM/c^2$. The dimensionless form of equation (\ref{d1})
is
\begin{equation}
\tilde{D}(r)={\frac {15\dot{m}(r)}{r}}\left[1-{\frac {\dot{m}_{\rm
in}}{\dot{m}(r)}}\left({\frac {r_{\rm in}}{r}}\right)^{1/2}\right],
\label{d2}
\end{equation}
where
\begin{displaymath}
\tilde{D}(r)=D(R){\frac {R^2\kappa_{\rm T}}{GMc}},~~~~\dot{m}={\frac
{\dot{M}}{\dot{M}_{\rm Edd}}},~~~~\dot{M}_{\rm Edd}={\frac {L_{\rm
Edd}}{0.1c^2}}
\end{displaymath}
\begin{equation}
L_{\rm Edd}={\frac {4\pi GMc}{\kappa_{\rm T}}},~~~~~~r={\frac
{R}{R_{\rm g}}},~~~~~~R_{\rm g}={\frac {GM}{c^2}}.
\end{equation}

In this work, we avoid being involved in the complexity of outflow
physics. In the region of the radiation flux being greater than the
maximal value given in equation (\ref{f_radmax1}), the radiation
force dominates over the vertical gravity for the gas in the upper
layer of the disc ($z/R>\tilde{H}_{\rm max}$), and it is blown away
by the radiation force. This makes the mass accretion rate decrease,
and therefore the energy dissipation rate decreases due to the
decrease of mass accretion rate in the disc. We assume that this
process continues till the radiation flux $f_{\rm rad}\simeq f_{\rm
rad}^{\rm max}$. We neglect the radial energy advection, and assume
$f_{\rm rad}(R)=D(R)$ in the disc, {which is to be justified later
in this section.}

\begin{figure}
\centerline{\psfig{figure=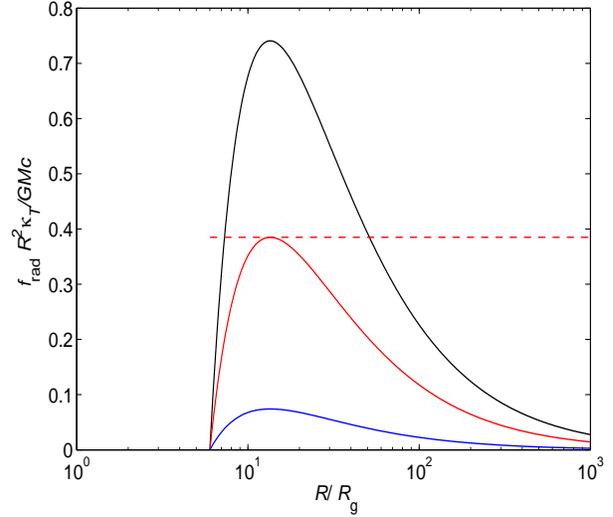,width=8.0cm,height=7.0cm}}
\caption{The radiation flux rates $f_{\rm rad}$ from the unit area
of the disc surface as functions of radius with different values of
$\dot{m}=\dot{m}_{\rm crit}=1.04$ (red), $0.2$ (blue) and $2$
(black). The red dashed line represents $f_{\rm rad}=f_{\rm
rad}^{\rm max}$. } \label{f_rad_r}
\end{figure}

There is a critical mass accretion rate, $\dot{m}_{\rm crit}$, below
which no outflow is driven by the radiation force. The critical rate
is calculated with
\begin{equation}
\tilde{D}(r)={\frac {15\dot{m_{\rm
crit}}(r)}{r}}\left[1-\left({\frac {r_{\rm
in}}{r}}\right)^{1/2}\right]=f_{\rm rad}^{\rm max},
\label{mdot_crit1}
\end{equation}
at $r=9r_{\rm in}/4$, because $\tilde{D}(r)$ reaches the maximal
value at this radius for a given $\dot{m}$. Using equation
(\ref{f_radmax1}), we obtain
\begin{equation}
\dot{m}_{\rm crit}={\frac {3\sqrt{3}}{5}}\simeq 1.04.
\label{mdot_crit2}
\end{equation}
The radiation flux $f_{\rm rad}(R)$ varying with mass accretion rate
$\dot{m}$ is plotted in Fig. \ref{f_rad_r}.


\textbf{}

\begin{figure}
\centerline{\psfig{figure=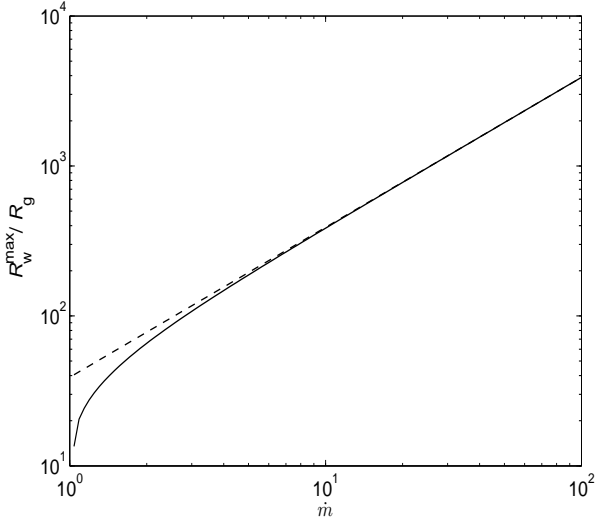,width=8.0cm,height=7.0cm}}
\caption{The outer radii of the region with outflows as functions of
mass accretion rate $\dot{m}_{\rm out}$. The solid line represents
the result calculated with equation (\ref{r_w_max1}), while the
dashed line corresponds to that calculated with equation
(\ref{r_w_max2}).   } \label{rw_mdot}
\end{figure}

For an accretion disc accreting at a rate $\dot{m}>\dot{m}_{\rm
crit}$, the outflows are present in the region of the disc between
$r_{\rm w}^{\rm min}$ and $r_{\rm w}^{\rm max}$. The gas at the disc
surface starts to be removed into the outflows from the disc at
$r_{\rm w}^{\rm max}$, where $\tilde{D}(r_{\rm w}^{\rm
max})=\tilde{f}_{\rm rad}^{\rm max}$. The outflows are switched off
at $r_{\rm w}^{\rm min}=9r_{\rm in}/4$, because $\tilde{D}(r)$ will
decrease within this radius (see Fig. \ref{f_rad_r}). This means
that the mass accretion rate at the inner radius of the disc
$\dot{m}_{\rm in}\equiv\dot{m}_{\rm crit}$ if the mass accretion
rate at the outer radius of the disc $\dot{m}_{\rm out}>\dot{m}_{\rm
crit}$. The outer radius $r_{\rm w}^{\rm max}$ of the disc with
outflows can be determined by solving the equation $\tilde{D}(r_{\rm
w}^{\rm max})=\tilde{f}_{\rm rad}^{\rm max}$, i.e.,
\begin{equation}
{\frac {15\dot{m}_{\rm out}}{r_{\rm w}^{\rm max}}}\left[1-{\frac
{\dot{m}_{\rm in}}{\dot{m}_{\rm out}}}\left({\frac {r_{\rm
in}}{r_{\rm w}^{\rm max}}}\right)^{1/2}\right]={\frac
{2\sqrt{3}}{9}}, \label{r_w_max1}
\end{equation}
with a given value of $\dot{m}_{\rm out}$. Using equation
(\ref{r_w_max1}), we have
\begin{equation}
r_{\rm w}^{\rm max}\approx{\frac {45\sqrt{3}}{2}}\dot{m}_{\rm out},
\label{r_w_max2}
\end{equation}
for the case of $\dot{m}_{\rm out}\gg\dot{m}_{\rm crit}$. The outer
radii of the region with outflows varying with mass accretion rate
$\dot{m}_{\rm out}$ calculated with equations (\ref{r_w_max1}) and
(\ref{r_w_max2}) are plotted in Fig. \ref{rw_mdot}. It is found that
the results almost converge when mass accretion rate $\dot{m}_{\rm
out}\ga 5$.


\begin{figure}
\centerline{\psfig{figure=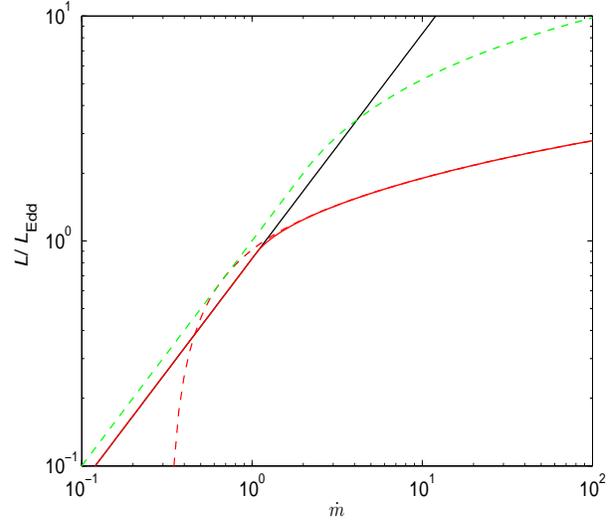,width=8.0cm,height=7.0cm}}
\caption{The relation of Eddington ratio $\lambda$ of the disc with
the mass accretion rate. The red solid line represent the result
calculated with equation (\ref{lum2}), while the red dashed line is
the result calculated with the analytical approximation
(\ref{lum3}). The black line indicates the results with standard
disc model, i.e., the Eddington ratio varies with the mass accretion
rate linearly. The green line represents the result of the slim disc
without outflows \citep*[see equation 19 in][]{2000PASJ...52..133W}.
The definition of $\dot{m}$ in their work is different from ours,
and their $\dot{m}$ has been divided by ten to be consistent with
our definition of $\dot{m}$.} \label{lum_mdot}
\end{figure}

The luminosity of an accretion disc with outflows
($\dot{m}>\dot{m}_{\rm crit}$) can be calculated with
\begin{equation}
{\frac {L}{L_{\rm Edd}}}={\frac {1}{L_{\rm Edd}}}\int\limits_{R_{\rm
in}}^{\infty}4\pi RD(R){\rm d}R=\int\limits_{r_{\rm
in}}^{\infty}{\frac {\tilde{f}_{\rm rad}(r)}{r}}{\rm d}r,
\label{lum1}
\end{equation}
where
\begin{equation}
\tilde{f}_{\rm rad}(r)=\left\{ \begin{array}{ll} {\frac
{15\dot{m}_{\rm in}}{r}}\left[1-\left({\frac {r_{\rm
in}}{r}}\right)^{1/2}\right],~~~~~~~~~~~~~~r\le r_{\rm w}^{\rm min};\\
\tilde{f}_{\rm rad}^{\rm max},~~~~~~~~~~~~~~~~~~~~~~~~~~~~~r_{\rm w}^{\rm min}<r<r_{\rm w}^{\rm max};\\
{\frac {15\dot{m}_{\rm out}}{r}}\left[1-{\frac {\dot{m}_{\rm
in}}{\dot{m}_{\rm out}}}\left({\frac {r_{\rm
in}}{r}}\right)^{1/2}\right],~~~r\ge r_{\rm w}^{\rm max}.\\
\end{array} \right.
\label{f_rad3}
\end{equation}
The integral in equation (\ref{lum1}) can be carried out
analytically as
\begin{displaymath}
\lambda={\frac {L}{L_{\rm Edd}}}={\frac {35\dot{m}_{\rm
in}}{27r_{\rm in}}}+{\frac {15\dot{m}_{\rm out}}{r_{\rm w}^{\rm
max}}}\left[1-{\frac {2\dot{m}_{\rm in}}{3\dot{m}_{\rm
out}}}\left({\frac {r_{\rm in}}{r_{\rm w}^{\rm
max}}}\right)^{1/2}\right]
\end{displaymath}
\begin{equation}
~~~~~~~~~~~~~~~~~~~~~~~+{\frac {2\sqrt{3}}{9}}\ln\left({\frac
{r_{\rm w}^{\rm max}}{r_{\rm w}^{\rm min}}}\right).
 \label{lum2}
\end{equation}
Substituting equation (\ref{r_w_max2}) into (\ref{lum2}), we obtain
an approximation for $\lambda$,
\begin{displaymath}
\lambda\approx{\frac {35}{162}}\dot{m}_{\rm in}+{\frac
{2\sqrt{3}}{9}}\left[1-{\frac {2\dot{m}_{\rm in}}{3\dot{m}_{\rm
out}}}\left({\frac {4\sqrt{3}}{45\dot{m}_{\rm
out}}}\right)^{1/2}\right] \end{displaymath}
\begin{displaymath}
~~~~~~~~~~~+{\frac {2\sqrt{3}}{9}}\ln{\frac
{5\sqrt{3}}{3}}\dot{m}_{\rm out}
\end{displaymath}
\begin{equation}
={\frac {57\sqrt{3}}{162}}-{\frac {8\sqrt{15}}{675}}\left({\frac
{\sqrt{3}}{\dot{m}_{\rm out}}}\right)^{3/2}+{\frac
{2\sqrt{3}}{9}}\ln{\frac {5\sqrt{3}}{3}}\dot{m}_{\rm out},
\label{lum3}
\end{equation}
in high-$\dot{m}_{\rm out}$ cases. We plot the Eddington ratio of
the disc changing with mass accretion rate $\dot{m}_{\rm out}$ in
Fig. \ref{lum_mdot}, in which the analytical approximation
(\ref{lum3}) agrees with the numerical result quite well if
$\dot{m}_{\rm out}\ga 1.5$. It is found that the Eddington ratio
$\lambda\propto\ln\dot{m}_{\rm out}$ when $\dot{m}_{\rm out}$ is
high.


\begin{figure}
\centerline{\psfig{figure=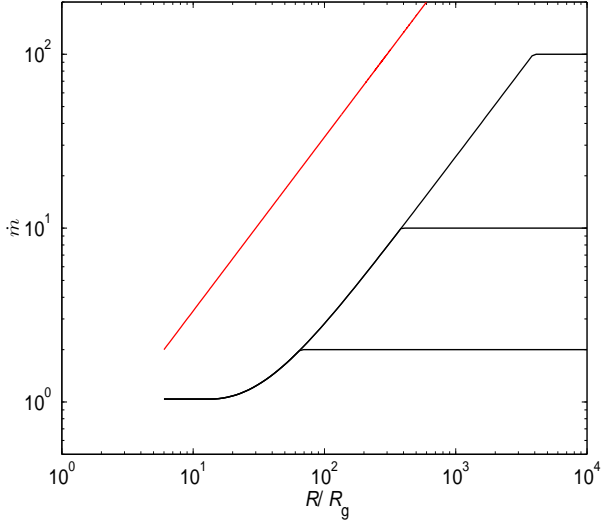,width=8.0cm,height=7.0cm}}
\caption{The mass accretion rates $\dot{m}(r)$ as functions of
radius with different values of $\dot{m}_{\rm out}=2$, $10$ and
$100$, respectively. {The red line is the outer radius of the
advection dominated region in the disc (see equation
\ref{r_adv1})}.} \label{mdot_r}
\end{figure}

As discussed above, the outflows are present when accretion rate is
higher than a critical value, which means that the mass accretion
rate is a function of radius. The outflows are driven from the disc
region between $r_{\rm w}^{\rm min}$ and $r_{\rm w}^{\rm max}$, and
the mass accretion rate in this region can be calculated by
\begin{equation}
\tilde{D}(r)={\frac {15\dot{m}(r)}{r}}\left[1-{\frac {\dot{m}_{\rm
in}}{\dot{m}(r)}}\left({\frac {r_{\rm
in}}{r}}\right)^{1/2}\right]=\tilde{f}_{\rm rad}^{\rm max},
\label{mdot1}
\end{equation}
which reduces to
\begin{equation}
\dot{m}(r)={\frac {\tilde{f}_{\rm rad}^{\rm max}}{15}}r+\dot{m}_{\rm
in}\left({\frac {r_{\rm in}}{r}}\right)^{1/2}={\frac
{2\sqrt{3}}{135}}r+{\frac {9\sqrt{2}}{5}}r^{-1/2}. \label{mdot2}
\end{equation}
The mass accretion rate remains constant radially, i.e.,
$\dot{m}=\dot{m}_{\rm out}$ in the region of $r\ge r_{\rm w}^{\rm
max}$, and $\dot{m}=\dot{m}_{\rm in}=3\sqrt{3}/5$ for $r\le r_{\rm
w}^{\rm min}\equiv9r_{\rm in}/4=27/2$. In Fig. \ref{mdot_r}, we plot
the $r$-dependent mass accretion rate $\dot{m}(r)$ with different
values of $\dot{m}_{\rm out}$. The mass accretion rate in the inner
edge of the discs remains constant, $\dot{m}_{\rm in}=1.04$, due to
the presence of outflows.

{\citet{2000PASJ...52..133W}'s calculations on the slim discs show
that radial energy advection is important in the inner region of the
discs, and the region of the disc with
\begin{equation}
r\le r_{\rm adv}\simeq 3\dot{m}, \label{r_adv1}
\end{equation}
is advection dominated \citep*[see equation 16 in][and note that
their $\dot{m}$ should be divided by ten to be consistent with our
definition of $\dot{m}$]{2000PASJ...52..133W}. The advection
dominated region disappears in the disc when $\dot{m}\la 2$, which
means that the advection is neglibile in this case \citep*[see
equation 19 in][]{2000PASJ...52..133W}. In Fig. \ref{mdot_r}, we
find that the mass accretion rate $\dot{m}$ derived in this work is
always much lower than that required for advection dominated case at
any radius in the disc. This justifies the assumption of
$D(R)=f_{\rm rad}$ adopted in our calculations.}


\begin{figure}
\centerline{\psfig{figure=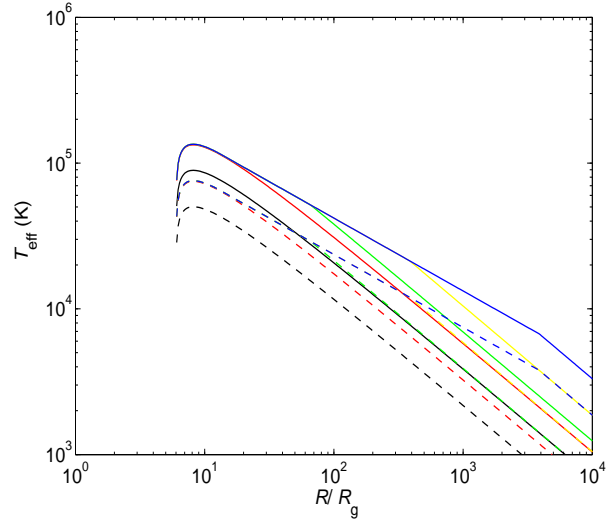,width=8.0cm,height=7.0cm}}
\caption{The effective temperatures of the discs as functions of
radius with different values of $\dot{m}_{\rm out}=0.2$ (black), $1$
(red), $2$ (green), $10$ (yellow) and $100$ (blue), respectively.
The solid lines indicates the results calculated with black hole
mass $M=10^8M_{\odot}$, while the dashed lines are for
$M=10^9M_{\odot}$. } \label{temp_r}
\end{figure}

\begin{figure}
\centerline{\psfig{figure=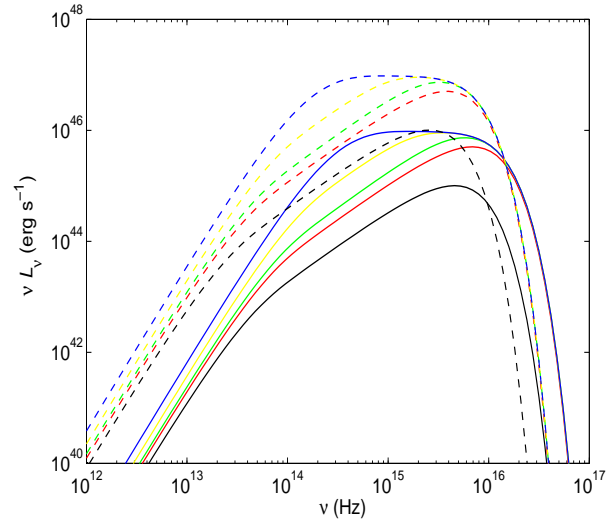,width=8.0cm,height=7.0cm}}
\caption{The spectra of the discs surrounding a black hole as
functions of radius with different values of $\dot{m}_{\rm out}=0.2$
(black), $1$ (red), $2$ (green), $10$ (yellow) and $100$ (blue),
respectively. The solid lines indicates the results calculated with
black hole mass $M=10^8M_{\odot}$, while the dashed lines are for
$M=10^9M_{\odot}$. } \label{spect_mdot}
\end{figure}

The effective temperature of the disc as a function of radius can be
calculated with $T_{\rm s}(R)=[f_{\rm rad}(R)/\sigma]^{1/4}$ by
using the derived radiation flux (see equation \ref{f_rad3}), when
the black hole mass $M$ is specified. The spectrum of the disc is
therefore available with the derived effective temperature of the
disc. We plot the effective temperatures of the discs surrounding
massive black holes with different values of mass accretion rate
$\dot{m}_{\rm out}$ in Fig. \ref{temp_r}. The temperatures of the
inner regions of the discs are almost the same for any values of
$\dot{m}_{\rm out}>\dot{m}_{\rm crit}$. In Fig. \ref{spect_mdot}, we
plot the radiation spectra of the accretion discs varying with mass
accretion rates. The spectra of the discs surrounding massive black
holes are saturated in UV/soft X-ray wavebands ($\nu\ga 10^{16}~{\rm
Hz}$) when accretion rate is as high as $\dot{m}_{\rm out}\sim 2$.


\section{Summary and discussion}\label{discussion}

\subsection{Structure of the disc}

The vertical component gravity increases with $z$, and reaches the
maximum at $z/R=\sqrt{2}/2$ (see Fig. \ref{f_grav_z}). {The
radiation force will dominate over the vertical gravity in the disc
if the radiation flux is sufficiently large. It implies that
outflows can be driven from the disc if the radiation flux of the
disc is greater than a critical value.} This has already been
discussed in the previous works
\citep*[][]{2006ApJ...652..518M,2007ApJ...660..541G}.

The relative disc thickness of $H/R$ corresponds to a dimensionless
radiative flux $\tilde{f}_{\rm rad}$ at the disc surface for a
radiation pressure dominated disc. We derive the critical mass
accretion rate $\dot{m}_{\rm crit}=3\sqrt{3}/5\simeq 1.04$, below
which the relative thickness $H/R$ of the disc is always less than
$\sqrt{2}/2$, i.e., no outflow is present (see Fig. \ref{f_rad_r}).
In the case of $\dot{m}>\dot{m}_{\rm crit}$, the outflows are driven
from the disc surface in the disc region between $r_{\rm w}^{\rm
min}$ and $r_{\rm w}^{\rm max}$, where $r_{\rm w}^{\rm min}\equiv
9r_{\rm in}/4$, and $r_{\rm w}^{\rm max}$ is a function of mass
accretion rate $\dot{m}$ (see equation \ref{r_w_max1}). The outer
radius of the disc region with outflows $r_{\rm w}^{\rm max}\propto
\dot{m}$ when $\dot{m}$ is large (see equation \ref{r_w_max2}). For
a disc accreting at an extreme rate $\dot{m}\sim 100$, the disc
region with outflows can extend to thousands of gravitational radii
(see Fig. \ref{rw_mdot}). Due to the presence of outflows, the mass
accretion rate is not a constant radially in the disc. A fraction of
gas at the disc surface is blown away by the radiation force, so we
assume that mass accretion rate is self-adjusted to let the
dissipation power in unit surface area of the disc $D(R)=f_{\rm
rad}^{\rm max}$. {This means that we have assumed the radial inflow
timescale at a certain radius to be longer than the timescale for an
outflow to change the accretion rate of the whole disc. The outflow
timescale can be estimated only if the dynamics of the outflow is
well studied, which is beyond the scope of this work.} The mass
accretion rate as a function of radius is derived in this work (see
equation \ref{mdot2}, and Fig. \ref{mdot_r}). At large radii, the
mass accretion rate $\dot{m}\propto r$, which is the same as the
result given in \citet{2007ApJ...660..541G}. This is qualitatively
consistent with the hydrodynamical simulations on super-critical
accretion flows \citep*[see, e.g., Fig. 6
in][]{2005ApJ...628..368O}. {Their numerical simulations indicate
that the outflow can be so strong to carried sufficient gas from the
disc, though the detailed study of the outflow dynamics is needed to
clarify this issue.}

{We note that the relative disc thickness $H/R$ is larger than that
derived in the previous works for the same mass accretion rate
$\dot{m}$ \citep*[e.g.,][]{2003ApJ...593...69K}. In most of the
previous works, an approximation of the vertical component of
gravity, $z\Omega_{\rm K}^2$, is adopted, which over-estimates the
vertical gravity compared with the exact one used in this work when
$z/R$ is large \citep*[see Fig. 1 in][]{2007ApJ...660..541G}. This
approximation leads to underestimate of $H/R$ in the previous
works.} For relative disc thickness $H/R\la 0.2$, the vertical
gravity $z\Omega_{\rm K}^2$ is a good approximation (see Fig.
\ref{f_grav_z}). {Our calculations show that the disc thickness $H$
is systematically larger than that derived in
\citet{2003ApJ...593...69K}. The relative disc thickness
$H/R\rightarrow \sqrt{2}/2$ in the disc with $R\la 400 R_{\rm g}$
when $\dot{m}=10$ (see Fig. \ref{mdot_r}), which is about twice of
that given in \citet{2003ApJ...593...69K} for the same accretion
rate (see Fig. 5 in that paper). In that work, the scale height of
the disc is estimated with
\begin{equation}
H^\prime={\frac {c_{\rm s}}{\Omega_{\rm K}}},
\end{equation}
where the sound speed at the mid-plane of the disc $c_{\rm
s}=(p_{\rm mid}/\rho_{\rm mid})^{1/2}$ , which is different from
ours even in the limit of $H\ll R$. In this work, the disc thickness
can be derived from equation (\ref{f_rad1}),
\begin{equation}
H\simeq{\frac {f_{\rm rad}\kappa_{\rm T}}{\Omega_{\rm
K}^2c}},\label{h_final}
\end{equation}
when $H/R\ll 1$. The outgoing radiation flux is
\begin{equation}
f_{\rm rad}={\frac {4acT_{\rm mid}^4}{3\kappa_{\rm T}\rho_{\rm
mid}H}}.\label{f_rad_t}
\end{equation}
Substitute equation (\ref{f_rad_t}) into (\ref{h_final}), we have
\begin{equation}
H\simeq{\frac {(4p_{\rm mid})^{1/2}}{\rho_{\rm mid}^{1/2}\Omega_{\rm
K}}}={\frac {2c_{\rm s}}{\Omega_{\rm K}}}=2H^\prime,\label{h_final2}
\end{equation}
where $p_{\rm mid}=aT_{\rm mid}^4/3$ is used for a radiation
pressure dominated disc. This can explain the systematic difference
of disc thickness between \citet{2003ApJ...593...69K}'s paper and
this work. }

{Our calculations show that outflows can be driven from the outer
region of the disc, which may be truncated at $R_{\rm out}$ where
the disc becomes gravitationally unstable
\citep*[e.g.,][]{1998A&A...337..625H,2004A&A...415...47K}. The outer
radius of the disc, $R_{\rm out}$, is a function of black hole mass,
accretion rate, and the viscosity parameter $\alpha$, which can be
calculated with derived disc structure. However, the detailed disc
structure has not been derived in this work, and the constraint on
the size of a disc with outflows will be reported in our future
work.} {Our present work cannot tell whether the outflow is damped
by the vertical self-gravity or not.}

\subsection{Luminosity of the disc}

The luminosity of the disc with outflows can be derived with
equation (\ref{lum2}). We find that, $L/L_{\rm
Edd}\propto\ln\dot{m}$, if $\dot{m}$ is large (equation \ref{lum3}).
This looks similar to the result for the slim disc
\citep*[][]{2000PASJ...52..133W,2006ApJ...648..523W}, however, the
physics is completely different. In their works, no outflow is
considered, and the mass accretion rate remains constant radially in
the disc.

The cooling timescale is longer than the viscous timescale of the
slim disc. Therefore, only a fraction of the dissipated power is
radiated away, and the remainder is advected inwards. The radiation
of the slim disc is limited by the photon trapping effect within a
certain radius, which is roughly proportional to mass accretion rate
$\dot{m}$. Thus, the luminosity of the slim disc $L/L_{\rm
Edd}\propto\ln\dot{m}$
\citep*[][]{2000PASJ...52..133W,2006ApJ...648..523W}. In our present
work, the mass accretion rate at the inner radius of the disc
$\dot{m}_{\rm in}=\dot{m}_{\rm crit}\simeq 1.04$ if $\dot{m}_{\rm
out}>\dot{m}_{\rm crit}$, because most of the gas in the disc is
carried away in the outflows. The luminosity of the disc is
therefore limited by the outflows (see Sect. \ref{disc_outfl}). We
find that the luminosity of the disc with outflows $L/L_{\rm Edd}$
increases more slowly than that for the slim disc (see Fig.
\ref{lum_mdot}). In this work, the disc luminosity $L/L_{\rm
Edd}\sim 3$ provided $\dot{m}\sim 100$, which is lower than
$L/L_{\rm Edd}\sim 10$ for the same accretion rate in
\citet{2000PASJ...52..133W}. {Our results are comparable with
$L/L_{\rm Edd}\sim 2.6$ for $\dot{m}\sim 100$, as given in
\citet{2003ApJ...593...69K}.}

{Our calculations are carried out based on the balance between the
radiation force and the gravity in the vertical direction of the
disc, which means that the main conclusions on the maximal disc
luminosity in this work will not be altered whether the energy
advection is important or not. We find that the radial energy
advection is always negligible in an accretion disc with outflows
(see Fig. \ref{mdot_r}, and the detailed discussion in Sect.
\ref{disc_outfl}).}

{The UV/soft X-ray spectrum of the disc surrounding a massive black
hole is determined by the effective temperature profile and electron
scattering effects \citep*[][]{2003ApJ...593...69K}. In this work,
the spectral calculations are carried out based on the local
blackbody assumption without including the electron scattering
effects. Therefore, the spectral shape of the disc in the high
frequency end} is predominantly determined by the structure of the
inner region of the disc. In Fig. \ref{mdot_r}, we find that the
structure of inner region ($\la 100 R_{\rm g}$) of the disc remains
unchanged when $\dot{m}_{\rm out}\ga 2$, from which most of the
accretion power is radiated. The effective temperatures of the discs
with outflows in the inner regions almost converge for different
values of $\dot{m}_{\rm out}$ (see Fig. \ref{temp_r}). Therefore, it
is found that the emission in the high frequency end of the spectrum
($\nu\ga 10^{16}~{\rm Hz}$) is saturated when $\dot{m}_{\rm out}\ga
2$ (see Fig. \ref{spect_mdot}). The further calculations of the
spectrum {including the electron scattering effects, as done by
\citet{2003ApJ...593...69K}, are necessary for detailed
astrophysical applications.} {As the structure of the inner region
of the disc with outflows accreting at a high rate is insensitive to
the mass accretion rate at the outer radius, we believe that the
feature of the emission saturation should still be present, though
the final results may be quantitatively different from our present
results.} Such calculations are beyond the scope of this work, which
will be reported in our future work.


\subsection{Implications of the model}

The saturation of the emission in the high frequency end of the disc
spectra can be used for estimating the masses of the black holes
accreting at high rates, such as, narrow-line Seyfert galaxies. It
implies that luminous quasars with $\dot{m}_{\rm out}\ga 2$ can be
used as a type of ``cosmological candles", and one does not need to
search massive black holes accreting at extremely high rates
\citep*[][]{2013PhRvL.110h1301W}.

The relation of $\dot{m}_{\rm out}$-$L/L_{\rm Edd}$ derived in this
work shows that the radiation of the disc can only be several times
of Eddington luminosity even if the mass accretion rate
$\dot{m}_{\rm out}\gg 1$. This seems to be in contradiction with the
observations of some ultra-luminous X-ray (ULX) sources, of which
the luminosity can be highly super Eddington if normal stellar mass
black holes reside in the ULX \citep*[e.g.,][and the references
therein]{2011NewAR..55..166F,2013MNRAS.435.1758S,2014Sci...343.1330S}.
As suggested by \citet{2001ApJ...552L.109K}, this issue can be
alleviated if the X-ray emission from the ULXs is mildly beamed to
the observer, and the observed X-ray emission is predominantly from
the outflows \citep*[][]{2006MNRAS.370..399B}, which are
qualitatively consistent with our results that outflows are
inevitably driven from the disc surface when $\dot{m}_{\rm
out}>\dot{m}_{\rm crit}$. The detailed investigation on such
outflows from the radiation dominated accretion disc will be
reported in our future work, which is beyond the scope of this
paper. An alternative possibility is that these ULX may have
intermediate mass black holes, of which the luminosity is therefore
lower or close to Eddington luminosity \citep*[see][for a review,
and references therein]{2011NewAR..55..166F}.

Luminous quasars are discovered at redshifts $z\ga 6$
\citep*[see][and the references
therein]{2003AJ....125.1649F,2011Natur.474..616M,2013ApJ...779...24V},
corresponding to the age of the universe less than 1~Gyr. To grow a
black hole to $\sim 10^9{\rm M}_\odot$ in such a short period of
time, the massive black holes in these quasars must have either
accreted persistently at a rate close to the Eddington value
\citep*[e.g.,][]{2005ApJ...620...59S,2009ApJ...696.1798T}, or
undergone many episode super-Eddington accretion events
\citep*[see][and the references
therein]{2003ApJ...596L..27K,2004A&A...420L..23K,2005ApJ...633..624V,2014arXiv1401.3513V}.
In this work, we find that the rate of mass accreted by the black
hole $\dot{m}_{\rm in}\sim 1$, even if the mass accretion rate at
the outer radius of the disc $\dot{m}_{\rm out}\gg 1$, because most
of the gas in the disc is removed in the outflows from the disc. Our
results seems to be consistent with the continuous accretion
scenario for massive black hole growth at high redshifts
\citep*[e.g.,][]{2005ApJ...620...59S,2009ApJ...696.1798T,2014arXiv1406.3023T}.

\section*{Acknowledgments}
This work is supported by the NSFC (grants 11222328, 11173043,
11121062, and 11233006), the Strategic Priority Research Program
``the Emergence of Cosmological Structures" of the CAS (grant No.
XDB09000000), and Shanghai Municipality.

\end{document}